\documentclass{PoS}

\newcommand{\bs}{\begin{slide}}
\newcommand{\es}{\end{slide}}
\newcommand{\bi}{\begin{itemize}}
\newcommand{\ei}{\end{itemize}}
\newcommand{\bq}{\begin{quote}}
\newcommand{\eq}{\end{quote}}
\newcommand{\be}{\begin{equation}}
\newcommand{\ee}{\end{equation}}
\newcommand{\bd}{\begin{displaymath}}
\newcommand{\ed}{\end{displaymath}}
\newcommand{\bea}{\begin{eqnarray*}}
\newcommand{\eea}{\end{eqnarray*}}

\newcommand{\la}{\langle}
\newcommand{\ra}{\rangle}

\title{Revisiting the pentaquark episode for lattice QCD}

\ShortTitle{Revisiting the pentaquark episode}

\author{\speaker{ASB Tariq}\thanks{ASBT would like to acknowledge the support of Lattice 2007 organisers and sponsors for funding the attendance at the symposium
and the Abdus Salam ICTP and SIDA for funding and arranging a visit to ASICTP where much of the work was done.  Discussions with Gregorio Herdoiza and comments from Jonathan Flynn and 
Chris Allton are also acknowledged.
}\\
        Department of Physics, Rajshahi University, Rajshahi 6205, Bangladesh\\
        E-mail: \email{asbtariq@ru.ac.bd}}

\abstract{This work revisits the pentaquark episode, particularly in reference to lattice QCD, to collect in one place the lessons that have been or should be learnt.  This 
also examines whether, and if yes, where, there might be any prospect for future studies of pentaquarks and the role of lattice QCD in it.
Tests such as volume dependence, and hybrid boundary conditions to distinguish bound and scattering states are identified as the positives to take forward from this episode.
On the other hand it is also observed that pentaquarks with heavier quark content and in the $SU(3)_f$ limit may still be worth not forgetting, and in that case lattice QCD 
may have an important role to play.}

\FullConference{The XXV International Symposium on Lattice Field Theory\\
		 July 30-4 August 2007\\
		 Regensburg, Germany}

\begin{document}

\section{Introduction}
The claimed pentaquark observations followed by seemingly conflicting sets of evidence from several experiments
and and similarly confusing conclusions in related theoretical studies that generated a good
deal of excitement for some time are now fading into the past.
However, this episode has left
behind interesting lessons in general, as well as for lattice QCD.
Here an attempt is made to collect in one
place, before we forget, the lessons learnt for lattice QCD.

\section{Background}

There is nothing in QCD forbidding tetra-, penta-, hexa-\ldots quark states.  
Manifestly exotic baryonic states, i.e.~states with quantum numbers that cannot 
be assigned to any single three-quark state, e.g.~ the $S=+1$ channel have 
therefore attracted a lot of attention. 
In fact, in the late 50's, even before the introduction of quarks, the $KN$ 
($K^+p$) system was being studied.  This got more extensive from the late 60's.
Then, these were known as $Z$-resonances or baryons.
By the mid 70's, things were heating up, with 
Particle Data Book 1976, devoting ten pages on the $S=+1$, $Z^*$
system:
\begin{quote}
{\it \ldots Three quarks cannot produce $S=+1$ baryon resonances $(Z^*)$'s, and this
has probably been the primary motivation for the great amount of experimental
effort that has gone into $S=+1$ baryon physics below $\sim$2 GeV/c during the
last several years.}
\end{quote}
This optimism, however, seemed to die out in the following decade with a three-page review
in Particle Data Book 1986, and then a damning verdict in the last review of the $Z^+$ 
system in Particle Data Book 1992:
\begin{quote}
{\it \ldots~the same story heard for
20 years \ldots The general prejudice against baryons not made of three quarks and the lack
of any experimental activity in this area make it likely that it will be another 20 years before the issue is decided.
Nothing new at all has been published in this area since our 1986 edition \ldots}
\end{quote}
The field looked killed, Particle Data Group stopped reviewing it and (nearly) everyone 
assumed that pentaquark states, if any, have very
large widths and if created {\it fall apart} into baryon-meson pairs.

However, there was some sort of revival of interest in late 90's with around thirty theory 
papers before 2000. In 1997, Diakonov, Petrov and Polyakov predicted a narrow width 
pentaquark state in the chiral soliton model \cite{Diakonov:1997mm}. 
In a remarkably bold and brief abstract, they write:
\begin{quote}
{\it We predict an exotic $Z^+$ baryon (having spin 1/2, isospin 0 and
strangeness +1) with a relatively
low mass of about 1530 MeV and total width of less than 15 MeV.  It seems that this region of masses has avoided
thorough searches in the past.}
\end{quote}
Perhaps, more importantly, Diakonov convinced Nakano to look for it at LEPS.  First observation 
of 1.54(1) GeV state with $\Gamma\leq 25$ MeV observed at
$4.6\sigma$ announced in October 2002 \cite{Nakano:2003qx}.  Ironically, the conference was titled
PANIC.

On the experimental side, after conflicting results 
from different experiments for a couple of years,
the negative experimental evidence started overriding the positive evidence.  
Most of the positive experiments have 
been contradicted either by the an upgraded version of the same experiment or 
an independent one at higher statistics.  
The few remaining
positive experiments are also in contradiction with negative experiments
as well as within positive experiments themselves, in particular, with the
spread of the peaks being too large to be statistically accounted for.
Further details are available in reviews e.g.~\cite{Danilov:2007bp}.
Overall, it can be said without much hesitation that, experimentally
the $\Theta^+$ and its partner pentaquark ($\Xi^{--}_5$,$\Theta_c$)
signals are all but dead.

\section{Position of lattice QCD}
Lattice QCD is the only way to do a non-perturbative QCD
calculation from first principles, i.e.~in a model-independent manner.
Lattice QCD does have some, not always well-appreciated, 
limitations as well as strengths.
Nevertheless, there was
a good deal of expectation placed on the lattice community.  This expectation
is portrayed in the following quotes from the following quotes from Jaffe and 
Wilczek \cite{Jaffe:2003sg} and Lipkin \cite{Lipkin} respectively
\begin{quote}
{\it ``On the theoretical side, one important direction is to bring
  the power of lattice gauge theory to bear on these issues.''} and \newline
{\it ``There is therefore interest both in experimental searches for
  the $P_{\overline{c}s}$ and in lattice gauge calculations.  The
  simplest lattice calculation with an infinitely heavy charmed
  antiquark and four light quarks uuds can easily be done in parallel
  \ldots both in the symmetry limit where all light quarks have the
  same mass and with SU(3) symmetry breaking''}
\eq
But what has been the response?  Against over five hundred, perhaps many more,
non-lattice papers, there were less than two dozens of lattice papers
from ten groups, most of which are listed \cite{Csikor,Sasaki:2003gi,
Chiu,Mathur:2004jr,Liu:2005yc,Ishii,Takahashi,Lasscock,Alexandrou:2005gc,
Holland:2005yt,Hagen:2006yi} and others omitted for brevity.

Initially there was almost some kind of frustation, as if more was
expected of the lattice community, prompting
remarks such as \cite{Burkert:2004rp,Close:2004ny}:
\begin{quote}
{\it "Lattice QCD is currently not providing fully satisfactory
predictions for the $\Theta^+$.  One group finds no signal, three groups find a signal at
about the right mass, two at negative parity, one at positive parity."} and \newline
{\it "It is time for Lattice QCD to recover some of its investment"}
\end{quote}
So what was the problem?

\subsection{Issues encountered in lattice calculations}

\bi
\item
Above strong decay threshold

Most important issue is that the $\Theta^+$ (1540
MeV) is above the $KN$ threshold (1433 MeV) The
$\Theta^+$ will be hidden in a tower of $KN$-scattering
states.  Fortunately on the lattice, scattering
states can only exist with discrete values of momentum --
so only a finite (usually one) number of $KN$-states can be in the
region of interest.  It needed to be understood whether any
states observed are $KN$ scattering states or the $\Theta^+$.
Not trivial to disentangle.

\item Scattering vs.~bound state

As opposed to single particle states, two-particle
ones have explicit $1/V$-volume dependence.  The 
Kentucky group \cite{Mathur:2004jr,Liu:2005yc} were the first to report a
very careful analysis addressing most issues, including this, rigorously and 
indeed they have observed the volume
dependence characteristic of $KN$ scattering states.

The Tsukuba group \cite{Ishii} has introduced a novel method using hybrid boundary
conditions (HBC). They have used different boundary conditions for different
flavours:
anti-periodic for $u,d$ and periodic for $s$. $K(u\bar s)$ and $N(udd)$
feel anti-periodic boundaries whereas $\Theta^+$ still has periodic
boundaries. $KN$ has non-zero momentum even in $s$-wave, lifting it above
the $\Theta^+$ threshold. Bound $\Theta^+(uudd\bar s)$ should experience no
shift.  But it shows $KN$ scattering state type behaviour.

In a comprehensive analysis with most suggested types of operators,
the Adelaide group has suggested using the mass splitting with $NK$ state
for analysis.
They consider the mass dependence of the splitting with a negative slope
being expected for bound states.
For all types of operators (with one questionable exception) they have observed
a positive slope leading to the conclusion that no signal of a bound state
is observed.

These three tests seem to be among the main gains for lattice QCD from this episode.
These are summarised in Fig.~\ref{fig:boundvsscatt}
\begin{figure}
\label{fig:boundvsscatt}
\includegraphics[width=4.7cm]{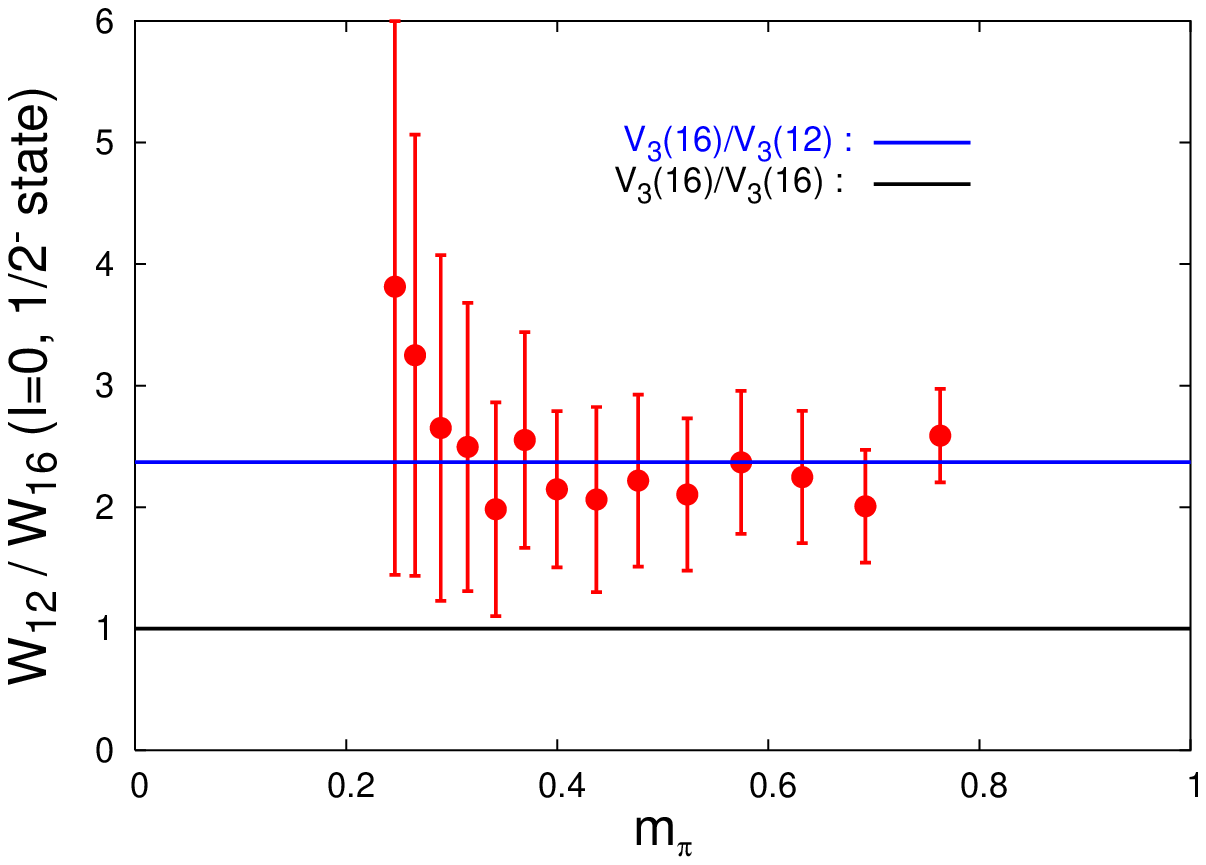}

\vspace{-3.9cm}

\hspace{5cm}
\includegraphics[angle=270,width=4.8cm]{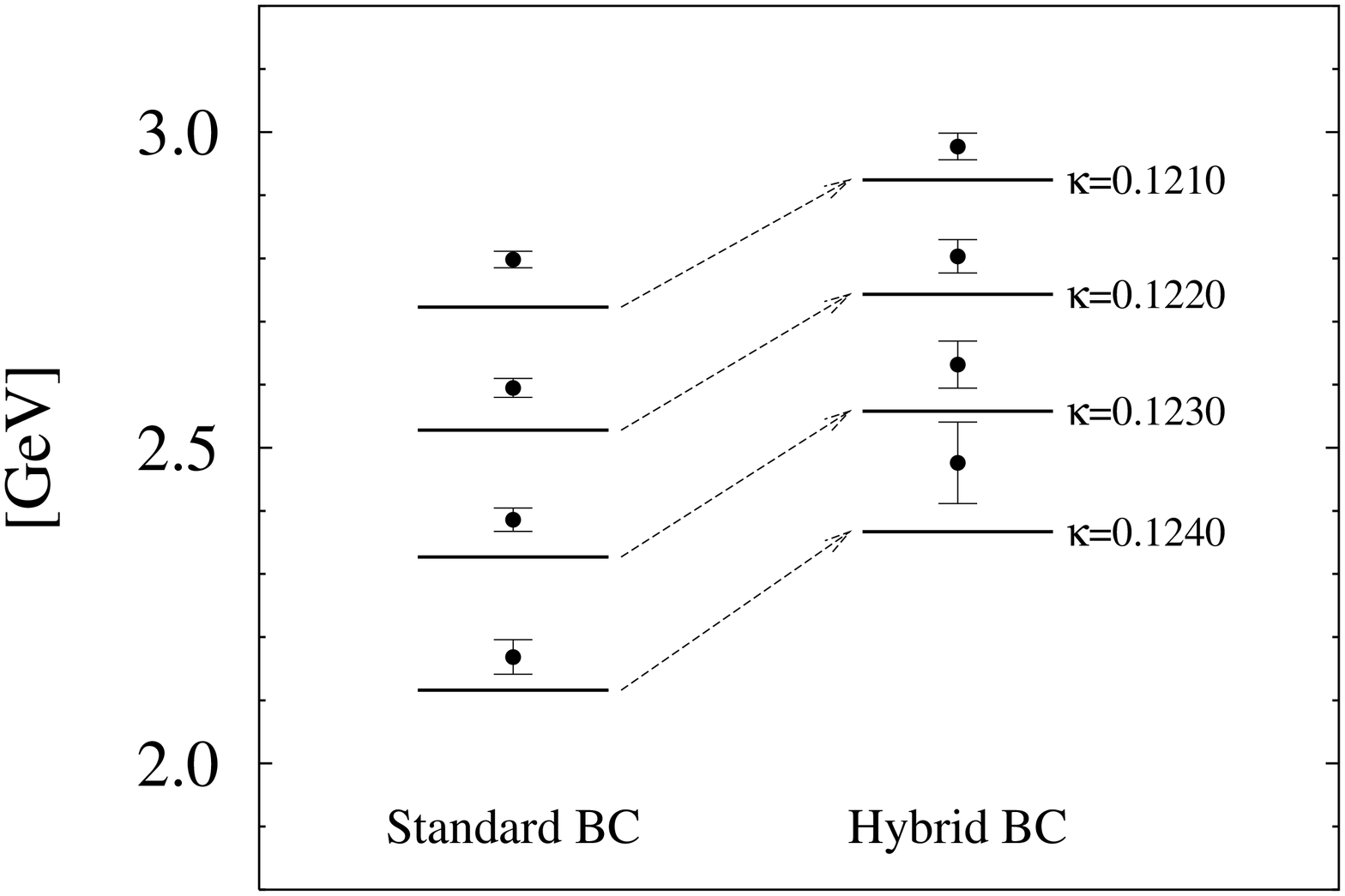}

\vspace{-3.1cm}

\hspace{10cm}
\includegraphics[angle=90,width=4.5cm]{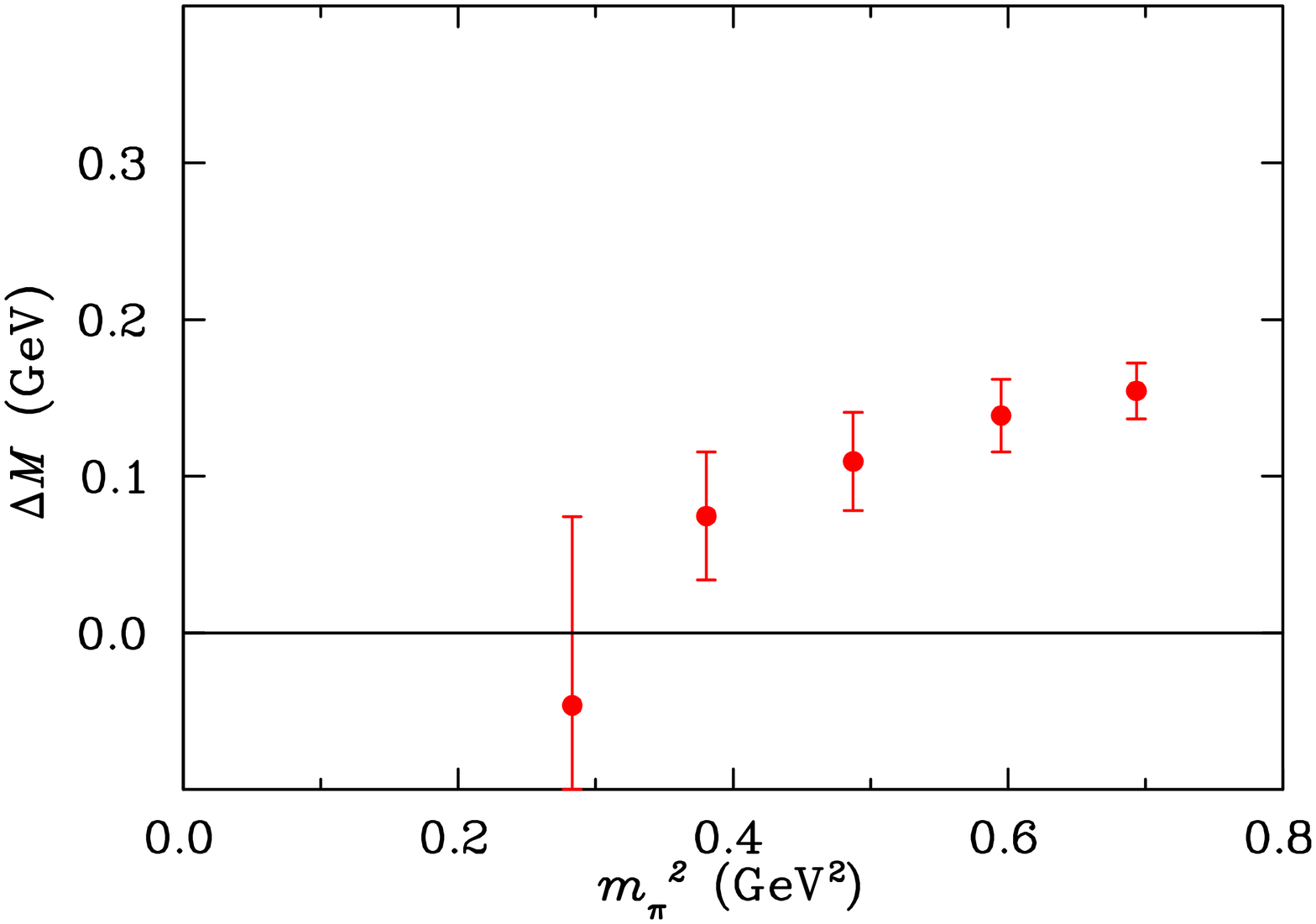}

\caption{Tests to distinguish bound states from scattering ones as have emerged 
from this episode: [left figure] $1/V$-type volume dependence characteristic of 
two-body system (no dependence expected for bound state)\cite{Mathur:2004jr}, [middle figure] states 
shifted for hybrid boundary condition and (no shift expected for bound state)\cite{Ishii} and
[right figure] mass splitting has a positive slope (negative expected for bound state)\cite{Lasscock} }
\end{figure}

\item What operator to use? 

A good deal of effort has been put into trying different operators. Cross
correlator approach seems to have emerged as the preferred one with the 
use of a set of interpolating operators.  Operators have been suggested 
such that $\la\Theta^+|{\cal O}_\Theta|0\ra\gg\la KN|{\cal O}_{KN}|0\ra$.
But with $\la KN|\Theta^+\ra\neq 0$ whatever
operator is used, the lowest state will be the $KN$.  This seems be 
confirmed from an important additional observation of \cite{Lasscock} 
(not stressed by the authors themselves) evident from Fig.~\ref{fig:lasscock2}
that the few different operators and actions produced similar masses
and many different calculations using different formalisms get very similar
masses for the states. Care, however, should be taken before generalising
this conclusion.
\begin{figure}
\begin{center}
\label{fig:lasscock2}
\includegraphics[angle=90,width=5cm]{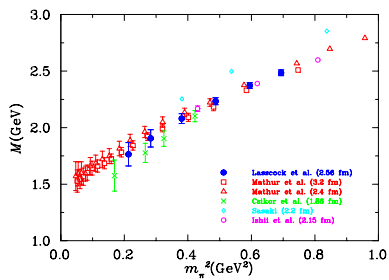}
\caption{Different operators and different actions leading to the same masses \cite{Lasscock}.}
\end{center}
\end{figure}
\ei

Briefly, what we have gained from this episode is a much better understanding of limitations of lattice simulations and issues related to simulation of states
above threshold, particularly, identification of bound and scattering
states using volume effects, hybrid boundary conditions, mass dependence of splitting etc.

If we accept what is seen in \cite{Takahashi,Alexandrou:2005gc} 
to be lattice artifacts or contamination from nearby states and the
analysis of \cite{Sasaki:2003gi} to be questionable, then it
seems we now have a consensus that whatever has been observed in lattice
studies are scattering states.

\section{Summary}

It is very intriguing to note how experiments and theoretical studies in
lattice QCD and sum rules followed very similar cirquitious paths, going
into misleading directions, but in the end reaching similar conclusions.
This exposed how much we scientists, despite trying to stress on objectivity,
are bound by human limitations of being prejudiced by what is going on around
us.  I have a long quote below from a sum rules
review \cite{Matheus:2006ix} of this episode only adding a "\/LQCD"(for 
lattice QCD) alongside the
QCDSR (for QCD sum rules).  Apart from saving me the trouble of writing 
this section myself, this illustrates strikingly how we all went through 
the same learning curve.

\bq
\it ``The first works on $\Theta^+$(1540) with QCDSR/LQCD addressed the
mass of the state and could all obtain a reasonable value of the
mass.  Later a more careful analysis revealed some problems with the
previous calculations.  In the mean time other pentaquarks were
observed: the $\Xi^{--}$ and the $\Theta_c$.  These were also studied
with QCDSR/LQCD. \ldots with more rigorous criteria it was more
difficult to reproduce the experimental data. \newline
``\ldots If, in the near future, the non-existence of pentaquarks is confirmed,
the community might address to the QCDSRLQCD practitioners the
following justified and embarassing question: ``how could you so
nicely calculate the correct mass of something that does not exist? \newline
``\ldots Looking back and taking distance, we might say that the work done 
over the last two years has undergone continuous improvements in
quality.  At the very beginning, in the heat of the discovery hours, some works
were done in rush and with a certain negligence in various
aspects. \ldots The second round of calculations went much
deeper in the details of QCDSR/LQCD procedures.  
However it was not a matter
of ``doing better'' what we already
knew how
to do. The method had to face new challenges. For example: in the
pentaquark study, for the first time, we were dealing
with a system that could be composed by independent subsystems,
like two non-interacting hadrons. It
has been a subject of debate how to disentangle and subtract
this component from the final results. Also, the more quarks
we have, the less unique is the definition of the interpolating
current. \ldots \newline
``To conclude we come back to the question raised in the introduction,
``How could we calculate the correct mass of something that does not
exist?''  In the light of the discussion presented
in the last sections, a sober answer would be: although
we started reproducing unfounded experimental results, it was
just a matter of time until we would reach a situation where,
reproducing these data would be so artificial as it was to use the
notion of ``aether'' in the years of the birth of special relativity.
At some point we would be obliged to push and twist the method so far,
that some more audacious groups would be brave enough to go against
the ``experimental evidence'' and put doubts on the experiments.  This
attitude was already taken by some phenomenologists, by some
experimentalists and by lattice theorists. \newline
``The efforts
of the community to overcome all these problems were very
productive.  All in all, we can say that pentaquarks have done
more for QCDSR/LQCD than these have done for pentaquarks. ''
\eq

\section{Outlook for pentaquarks}

As mentioned in the end of the last section, lattice QCD seems to have gained
more from than it has contributed to the episode.
It is time we do something to return the favour.  But what can
we do?

The recent observations not being sustained only means that there is no 
stable (narrow width) pentaquark state in this region. 
It does not at all imply that there are no pentaquarks, in fact still
nothing in QCD forbids such multiquark states.  It, however, is possible
that all such states simply fall apart.
There are indeed some interesting predictions: e.g.~(a) some quark models predict better binding for $\Theta_c$
and $\Theta_b$ (e.g., \cite{Jaffe:2003sg}), 
(b) some predict binding in the infinite $b$-quark mass
(static) limit (e.g., \cite{Oh:1994ky}),
(c) some predict better binding for $\overline{Q}qqqq$ when
one or two of the $q$s is a strange quark (e.g., \cite{Lipkin,Riska:1992qd}),
(d) relations e.g. $m(QQq)+m(qqq)\leq m(Qqq)+m(Qqq)$ suggest that $QQqqq$ could
be bound,
(e) some predict binding in the $SU(3)_f$ limit (e.g., \cite{Fleck:1989ff}) etc. 
What make it more interesting for lattice QCD are the facts that, (a) the $\Theta_b$ and $Qsqqq$, $QQqqq$ states are still
difficult experimentally, (b)the static and $SU(3)_f$ limits do not exist in experiments,
but do exist in lattice QCD.

It seems, here is another area where lattice QCD can go ahead of experiments.

\section{Conclusions}
Despite initial fumbles, the episode with the $\Theta^+$ has
enriched our understanding.  In particular we have learnt and/or refined
important techniques to separate scattering states from bound
states. Any future simulation of above/near-threshold states would 
have to satisfy these tests, e.g. 
\bi
\item volume effects
\item hybrid boundary conditions
\item mass dependence of splitting
\ei

And, there might still be interesting areas in pentaquark spectroscopy,
where lattice QCD can make valuable contributions, in fact, even
precede experiments, particularly for cases of 
\bi
\item one or more heavy/heavier quarks
\item static limit
\item $SU(3)_f$ limit.
\ei

\end{document}